\documentstyle[floats,aps]{revtex}
\input{psfig}
\begin{document}
\def \be{\begin{equation}}
\def \ee{\end{equation}}
\def \bea{\begin{eqnarray}}
\def \eea{\end{eqnarray}}
\def \bn{{\bf n}}
\def \br{{\bf r}}
\def \bk{{\bf k}}
\def \bQ{{\bf Q}}
\def \bm{{\bf m}}
\def \half{{1\over 2}}
\def \nd{{\vphantom{\dagger}}}
\def \cP{{\cal P}}
\def \cO{{\cal O}}
\twocolumn[\hsize\textwidth\columnwidth\hsize\csname @twocolumnfalse\endcsname

\title{Which tunnel faster across a quantum Hall strip: fractional charges or electrons?}
\author{Assa Auerbach}
\address{Department of Physics, Technion, Haifa 32000, Israel.}
\date{\today}
\maketitle
\begin{abstract}
The  tunneling rate $t_{1/3}$ of fractional charge across a $\nu=1/3$ Laughlin
state on the cylinder is computed numerically using Laughlin states on the cylinder. 
The decay with strip width $Y$ is fitted to
$t_{1/3}\propto \exp(-\alpha Y^2/
(12 \lambda^2))$ where $\lambda$ is the Landau length, and $\alpha\simeq 1.0$.
This rate is exponentially {\em larger} than the electron tunneling rate $t_1\propto \exp(-Y^2/(4  \lambda^2))$, and can be interpreted by analogy to a
superfluid vortex tunneling problem.
Experimental implications include the  ``law of corresponding states'',  periodicity of Aharonov-Bohm resistance oscillations and charge measurements by quantum shot noise.\\
pacs 73.40.Hm 
\end{abstract}
\vskip2pc]
\narrowtext
Transport of fractional
charges in quantum Hall systems has important experimental
manifestations.  Quasiparticles and 
edges of an  incompressible Hall liquid strip of bulk filling fraction $\nu=1/m$,
($m$ is an odd integer \cite{QHE}),
have quantized fractional charges whose values determine the following effects:  
(i)  The leading powers (of current and temperature)  of the longitudinal   voltage drop depend  on the tunneling charge $Q$, as shown by Wen using Luttinger liquid edge theory\cite{Wen}. 
(ii) The Aharonov\--Bohm (AB)  flux periodicity $\Delta\phi$
of the current oscillations  in a Cor\-bino disk is related to the tunneling charge 
by $Q$= $e\phi_0/\Delta\phi$\cite{GT}
(iii) The charges which dominate the backscattering current $I_B$ can be measured by the magnitude of the quantum shot noise $S$ by $S=2QI_B$ \cite{QSN-theory}.   Recent experiments in the $\nu=1/3$ phase, report 
excellent fits to fractional charge  $Q=e/3$ \cite{QSN-exp}.

Impurities allow tunneling of {\em both} fractional   charges and electrons across the bulk of a Hall fluid. The question of the title is therefore relevant to the observations of
experimental probes (i)-(iii),  since the latter depend
on the  {\em relative} rates of
fractional versus integer charge tunneling.
Kane and Fisher (KF) \cite{KF}  calculated the   renormalization group  
flows of  the tunneling coupling constants due to low lying   Luttinger liquid edge 
excitations. 
They found  that for
$\nu<1$,  electron tunneling becomes irrelevant at low enough temperatures while 
fractional  charge tunneling flows to strong coupling as the infra-red cut-off is reduced.  
In KF theory, however, the inter-edge scattering parameter
is an  undetermined parameter, which 
leaves the possibility that it might be undetectable at experimental temperatures.

This Letter presents a microscopic calculation
of  fractional  charge tunneling rates across a Hall fluid.  The numerical results show that
that the fractional charge tunneling rate is much larger than the electron charge rate at large strip widths. Subsequently, we connect the microscopic tunneling matrix
element to the inter-edge scattering parameter of KF theory, and discuss its
experimental implications.

Our domain is  the open cylinder 
$x\in[0,2\pi R)$, $-\infty < y < \infty$,   with
$N$   electrons, and a radially penetrating field $B={\phi_0\over 2\pi \lambda^2}$, where
$\lambda$, the Landau length, is henceforth our unit of distance.
This geometry  can describe
a quantum Hall liquid  strip with  two symmetric edges.

The free electron states   of the lowest Landau level (LLL)
are labelled by momenta  
$k=\gamma n$,  $n$ integer, and
$\gamma\equiv 1/R$. The wavefunctions are
\be 
\psi_k(x,y)=\sqrt{\gamma\over 2\pi  } \exp\left(ik x-{(y-k)^2\over 2}\right) .
\label{sp}
\ee 

\begin{figure}[htb]
\centerline{\psfig{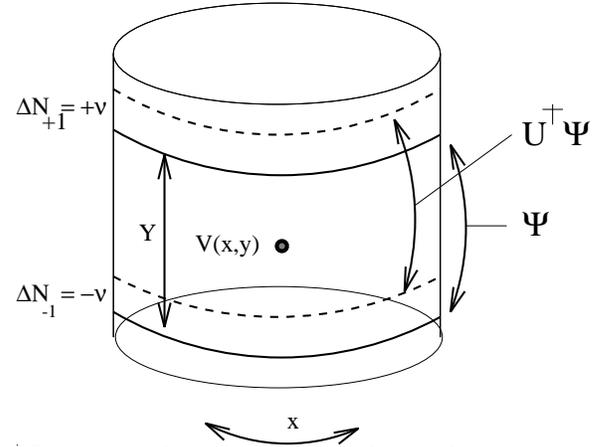}}
\caption{
\label{fig1}
Fractional charge tunneling depicted by two displaced Laughlin states of bulk density $\nu$ on the cylinder. $\Delta N_p$ are the edge charge differences, and $V$ is an
impurity potential which enables a transition between the states.
}
\end{figure}

The Laughlin state of filling fraction $\nu={1\over m}$ on the cylinder
was given by Thouless\cite{thouless}
\be
\Psi^{1/m}~= \prod_{i<j}\left(e^{ i \gamma(x_i+iy_i)} -e^{ i \gamma (x_j+iy_j)} \right)^{m}
\prod_i e^{-y_i^2/2}
\label{Psi-T}
\ee
It is the ground state of a suitably defined pseudopotential Hamiltonian
\cite{QHE,RH}. The expansion of  $\Psi^{1/m}$ in the 
LLL Fock basis is
\be
\Psi^{1/m}_L = \sum_{[\bk] } A[{\bf k}/\gamma] \exp\left( \sum_i k_i^2
\right) |\bk\rangle
\label{psi}
\ee
where $|\bk\rangle=|k_1,\ldots,k_N\rangle$ and $k_i\in \left[0,Y\right]$.  $Y=m\gamma (N-1)$ 
is  defined as the {\em width} of the Hall liquid strip (the width of the
area partially occupied by electrons depicted between the horizontal solid lines in Fig. \ref{fig1}.

There is an infinite family of other degenerate groundstates labelled by the total momentum $P=\sum_i k_i$, which are given by uniformly
shifting the momenta $k_i$ moving the electron density up or down the cylinder. 
A weak  ($v\!\to\! 0$) confining potential $V(y)={v\over 2} ( y-Y/2)^2$,   selects (\ref{psi}) as the ground state.

The expansion coefficients $A$ are given by \cite{jap}
\be
A[{\bf n}] = {1\over N!} \sum_{{\bf r}^1, \ldots {\bf r}^m }
(-1)^{\sum_{l=1}^m \cP[{\bf r}^l ] }
\prod_{i=1}^N \delta\left(n_i-\sum_{l=1}^m r_{i}^l  \right)
\ee
where  ${\bf r}^l$ is a permutation  of the set $0,1\ldots,N-1$, and $\cP$ is
the parity of a permutation. 

The  coefficients $A$  have a complicated structure\cite{itzik}, but it is useful to note
that the components with $A[\bk/\gamma]\ne 0$   can be derived from a single parent 
Tao-Thouless (TT)   state\cite{TT} 
\be
|\bk^{TT}\rangle  = |0, m\gamma ,2m\gamma, \ldots, Y\rangle 
\ee
For this state $A[\bk^{TT}/\gamma]=1$. All other $|\bk\rangle$  components 
are given by
successively  squeezing pairs of momenta toward each other. 
Rezayi and Haldane have shown \cite{RH} that in the regime $1<<Y << N$  the occupation number is constant for $k$ far from the edges, i.e. 
$n_k =\langle c^\dagger_k c^\nd_k\rangle =1/m$ for 
$ 1<k< Y-1$.

An impurity potential  in the LLL Fock representation is
\be
{\cal V} =\sum_{kk'}V_{k,k'} c^\dagger_k c^\nd_{k'}
\ee
where $c^\dagger_k$ creates an electron in state $\phi_{k}$.
The ground state to ground state tunneling rate of charge $qe$ between the 
edges, to  leading order  in
${\cal V}$, is 
\be
t_q= \langle \Psi|{\cal V}U^{m q } |\Psi\rangle, 
\label{tj}
\ee
where  $U$ is the unitary phase operator which translates all the single particle momenta by one interval
\be
U^\dagger  c^\dagger_k U  = c^\dagger_{k+\gamma}.
\label{U}
\ee  
$U^\dagger$  moves a fractional charge $1/m$  from  the $p\!=\!-1$ to the $p\!=\!+1$ edge
(see Fig. \ref{fig1}), and thus increases the total momentum of the Hall state by  
$P\to P+Q$, where  
$Q=N\gamma$.

\begin{figure}[htb]
\centerline{\psfig{figure=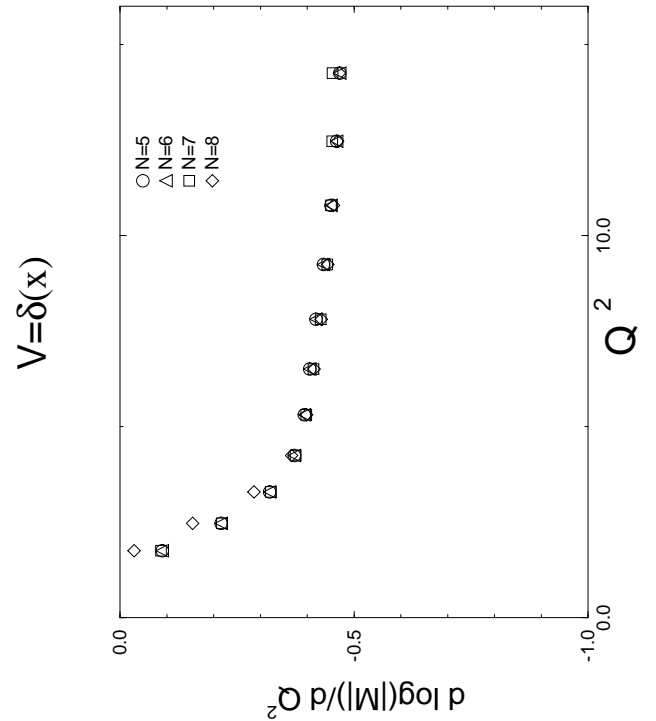,width=3.5in}}
\vspace{0.5in} 
\caption{
\label{fig2}
Numerical evaluation of the asymptotic decay of the many-body factor $M(Q)$, see Eq. (\protect\ref{MQ}) for the  $\nu=1/3$ Laughlin states. (a) and (b) show similar dependence
for two different localized impurity potentials. 
}
\end{figure}

For a weak impurity potential, the tunneling rate of a fractional charge 
is thus given by
\bea
t_{1/m}&=&\langle \Psi | {\cal V}U|\Psi \rangle
= \sum_{k}V_{k,k+Q} M_{k, k+Q} 
\nonumber\\
M_{k, k+Q }&=&  {1\over Z}\sum_{\bk,\bk'}  A[\bk/\gamma]A[\bk'/\gamma]  
e^{{1\over 2}(\bk^2+{\bk'}^2)} \nonumber\\
&&~~~~~ \times\prod_{n=1}^N \delta^N(\bk+\bQ(n),\bk'-\gamma {\bf 1}) 
\label{eq-M}
\eea
where  $Z\!=\! \langle \Psi |\Psi\rangle$,  $Q_i(n)\!=\!Q\delta_{in}$, and ${\bf 1}_i\!=\!1$.
$M_{k, k+Q }$  reflects the many body  overlap  of the relatively displaced Laughlin states. Its weighted sum  $M(Q)\!=\!V^{-1} \sum_{k}V_{k,k+Q}M_{k, k+Q }$
was computed numerically for local impurity potentials $V\delta(x)$ 
and  $V\delta(x)\delta(y-Y/2)$. The calculation was carried out for $\nu=1/3$ states with  5 up to 8 electrons.
As shown in  Figs. \ref{fig2} at large widths we find the asymptotic decay 
\be
 |M (Q)| \propto \exp(-{\alpha\over 2} Q^2)
\label{MQ}
\ee
where  $\alpha\approx 1.0$, and  independent of the number of electrons.
Combining  (\ref{MQ})  with  $V_{0,Q}\propto  \exp(-{1\over 4} Q^2)$  yields 
the tunneling rate's asymptotic dependence on  width
\be
t_{1/3} \sim   \exp\left( -\alpha Y^2/12\right)
\label{qpt}
\ee

In comparison, a unit charge tunneling rate, which is proportional to 
the potential matrix element, is
\be 
t_{1}=\langle \Psi|{\cal V}U^{3} |\Psi\rangle=n_{0}^2 V(0,Y) 
\ee 
For a localized potential of the form $V\delta(x)\delta(y-Y/2)$,  
\be
t_{1} \sim   \gamma^2 \exp\left( -Y^2/4\right)
\label{et}
\ee
where $n_{0 } \approx \gamma^\beta$ is appropriate for a density profile  which vanishes as a power law $n_k \approx k^{\beta}$ at the edge.  (The numerical results for $n_k$ 
in the Laughlin state (\ref{psi})  up to 8 particles is  $\beta=1.0$.) Thus, 
{\em the tunneling exponent  is  three times larger for quasiparticles
than for electrons.}

This result could be understood using the superfluid description of the 
 the fractional Hall phase,  which can be derived by
the Chern-Simons Ginzburg-Landau functional (CSGL) \cite{CSGL}. At the mean field level, the
ground state is a Bose  superfluid of  density 
$\rho_s ={1\over m} B/\phi_0$. The
dissipation of current involves tunneling of vortices between opposite edges, where
a vortex of unit circulation carries a  fractional electric charge of $e/m$.
Ignoring auxiliary gauge field fluctuations, and interactions at the core size, the vortex dynamics are governed by a Magnus force $e\phi_0\rho_s {\bf v} \times {\bf \hat z}$.
Thus they are quantized as  particles with  charge $e$ in the lowest Landau level of an effective field  ${\tilde B}=  \phi_0\rho_s$, and Landau length $\tilde \lambda = m \lambda$, with wave functions given by  (\ref{sp}). For $m=3$, the matrix element of
$\cal V$ between two vortex wavefunctions at the edges 
readily recovers  (\ref{qpt}), with $\alpha=1$.

{\em How do tunneling matrix elements couple to edge excitations?}
A half-strip density operator is defined as follows:
\bea
\rho_p(q)&=& \int_{Y/2}^{Y_p}dy \int_0^{2\pi R}dx e^{iqx} \rho(x,y)\nonumber\\
 &\approx&\sum_k \theta_{p,k}   \theta_{p,k+q}  c^\dagger_{k+q}c^\nd_{k} + \cO(q^2)
\label{rhopk}
\eea
where $p=\pm 1$,  $Y_p= (1+p)Y/2$ are the two edge $y$-coordinates,
and
\be
\theta_{pk} =\cases{1&$ p(k-Y/2)>0$\cr
0 &$p(k-Y/2)<0$}
\ee
The last  approximation in (\ref{rhopk})
applies to the long wavelength regime  $q<<1$.
The commutation relations of $\rho_p(q)$  are
\bea
\left[ \rho_p(q),\rho_{p'}(q')\right]&=& \delta_{pp'}  \delta_{q,-q'} 
\sum_k   \theta_{pk}   \theta_{p,k+q}  \left( n_{p,k} -n_{p,k+q}\right)
\nonumber\\
&&~~~~~~+  \delta_{pp'} \{q\ne -q'\} \nonumber\\
\label{comm}\eea
Since excitations in the bulk have an energy gap $ \Delta_B$,
the low energy sector 
includes only  particle-hole excitations  near the edges, i.e. 
$c^\dagger_k c^\nd_{k+q}\Psi$
with $k\approx Y_{p}$, and energies $\omega_q = vq$, where $v$ is the gradient of the confining potential.     $\{q\ne q'\}$ terms in  (\ref{comm})
create   excitations  deep in the bulk which introduce corrections suppressed by 
factors of $\omega_q/\Delta_B$ and $q/Y$. 
Also, in this sector $n_{p,k}$ is approximately diagonal
\be
n_{p,k}\simeq \cases{\nu& $ p(Y_p-k) >>1$\cr
0& $ (Y_p-k) >>1$}
\ee
Thus,  Wen's Kac-Moody algebra of edge bosons \cite{Wen} is recovered:
\be
 \left[ \rho_p(q),\rho_{p'}(q')\right]  \simeq \delta_{pp'}  \delta_{q,-q'}  \gamma^{-1}\nu q 
\label{km}
\ee

The edge charge operator is $N_p=\sum_k \theta_{p k} n_{p,k}$, which is conjugate to the edge phase operators $U_p$  
\be
\left[N_p, U_p^\dagger \right] = p n_{p,Y/2} U_p^\dagger \simeq  p\nu U_p^\dagger
\ee
The total phase operator (\ref{U}) is  $U^\dagger=U^\dagger_1 U_{-1}$.
The edge quasiparticle creation operator is constructed  following Haldane \cite{LL-Haldane}
\bea
\phi_p &=& p  \gamma \left(x   N_p/2 + i\sum_{q \ne 0} \theta(-pq) {e^{-iqx}\over q} \rho_p(q)\right)\nonumber\\
\psi^\dagger_p(x) &=& {\cal A}  e^{ i  Y_p x}e^{i \phi_{p}^\dagger(x)} U^\dagger_p  e^{ i \phi_{p}(x)}
\eea
where ${\cal A}$ is an  undetermined normalization constant. $\psi^\dagger_p(x)$ creates a localized edge excitation of  extra charge $\nu$ as evidenced by the commutator with $\rho_p(x)\! \equiv \! \sum_q e^{iqx} \rho_p(q)$: 
\be 
\left[\rho_p(x),\psi^\dagger_{p'}(x')\right]
=\nu \delta_{pp'}\delta(x-x') \psi^\dagger_{p'}(x) 
\ee 
The impurity potential operator in the low energy sector, simply
transfers a localized fractional charge between the edges. It must therefore be 
proportional to the normal ordered operator
\bea
 {\cal V}(x) &=&:\psi^\dagger_{p}(x) \psi_{-p}(x) :+ \mbox{H.c}\nonumber\\
&=&   {\cal A}^2 e^{iYx} e^{i\sum_p p \phi^\dagger_p(x)} U^\dagger  e^{i\sum_p p \phi^\nd_p(x)} + \mbox{H.c.}
\label{V-LL}
\eea
The normalization ${\cal A}^2 $ is precisely the 
bare fractional charge tunneling parameter
amplitude in KF theory \cite{KF}.
It can
now be determined by sandwiching both sides of Eq. (\ref{V-LL}) between the relatively displaced ground states  
leading to
\be
{\cal A}^2 =\langle \Psi |{\cal V}U|\Psi \rangle= t_{1/m}  
\ee

Two comments.  (i) Tao and Haldane \cite{TH}  have shown that in the absence of an  impurity potential, the quantum 
Hall ground state of $\nu=1/m$ on the {\em torus} has an $m$ fold degeneracy.  A  
time dependent AB flux threading the torus, moves the ground state between the $m$ different
states of this manifold, and the Hall conductance is precisely $\sigma_{xy}={1\over m}e^2/h$.
An impurity potential couples between degenerate ground states,
as it does on the cylinder, opening a minigap $\Delta$ between the ground state and the first excited state \cite{comm}. The  quantum Hall effect  can be observed provided the flux does not vary extremely slowly \cite{TH}, i.e. $\Delta/\hbar << \phi_0/V_x$, where $V_x$ is the induced electromotive force. 

(ii) For the infinite plane geometry, 
the tunneling exponent between two 
{\em localized} quasiparticle states centered on delta function impurities at
$\br_1,\br_2$  is\cite{lcs}  
\be
S(\br_1-\br_2) \propto \exp\left(- {B\over 4 m\phi_0 } |\br_1-\br_2|^2\right)
\ee
The scaling of the tunneling action with $B/m$ was argued to be a general 
property of low elementary excitations with  fractional charge $1/m$.
Generalizing this idea further, 
Jain, Kivelson and Trivedi \cite{JKT} have formulated a ``law of corresponding states'' which relate the dissipative response of  quantum Hall liquids at different filling fractions
using a conjecture that they scale with
${e\over m } B$ .  The results reported here are consistent with  this law, provided we
restrict ourselves to {\em wide} Hall strips in the presence of {\em weak} impurity potentials.

{\em Discussion}: KF have shown that edge excitations enhance the fractional charge and suppress
the unit charge 
contributions to the backscattering current\cite{KF}.  Thus the renormalization group flow enhances the bare tunneling ratio,  which strengthens the experimental
relevance of KF theory.

Current oscillations  in the Corbino disk in a slowly time dependent
AB flux can measure the dominant transport mechanism between edge states\cite{GT}.
Let us first consider {\em open leads} in the Hall voltage terminals.  
The impurity potential opens minigaps between the adiabatic energy curves as a function of AB flux $\phi_{AB}$.  
The lowest level crossing is at $\phi_{AB}\!=\!\phi_0$ between curves with minima separated by $\phi_0$,
with a minigap of size  $t_{1/3}$.   A minigap $t_1$ opens at higher energies
between
curves with minima separated by $3\phi_0$.
Since $t_{1/3}>>t_1$  we   expect to see only a periodicity of $\phi_0$ for open leads. If, on the other hand, we short the two Hall terminals with a low resistance wire, electron tunneling between the edges is enabled.
Thus, by varying the AB flux faster than the minigap $t_{1/3}$ we could observe a
periodicity of $3\phi_0$, reflecting transfer  of unit charges through the short.

Finally,  
the quasiparticle charge which dominates the backscattering
current between the edges can be measured by  quantum shot noise at zero temperature and bias \cite{QSN-theory}. Recent experimental  reports of measuring  fractional  charge in quantum shot noise of fractional quantum Hall systems\cite{QSN-exp}  are consistent with the expectation that fractional charges  
tunnel faster than electrons.

{\em Acknowledgments}:  The author gratefully acknowledges discussions with  E. Shimshoni,  D. Haldane  and M. Milovanovic. This work was supported by a grant from  Israel Science Foundation, and the  Fund for Promotion of Research at Technion.

\begin{figure}[htb]
\centerline{\psfig{figure=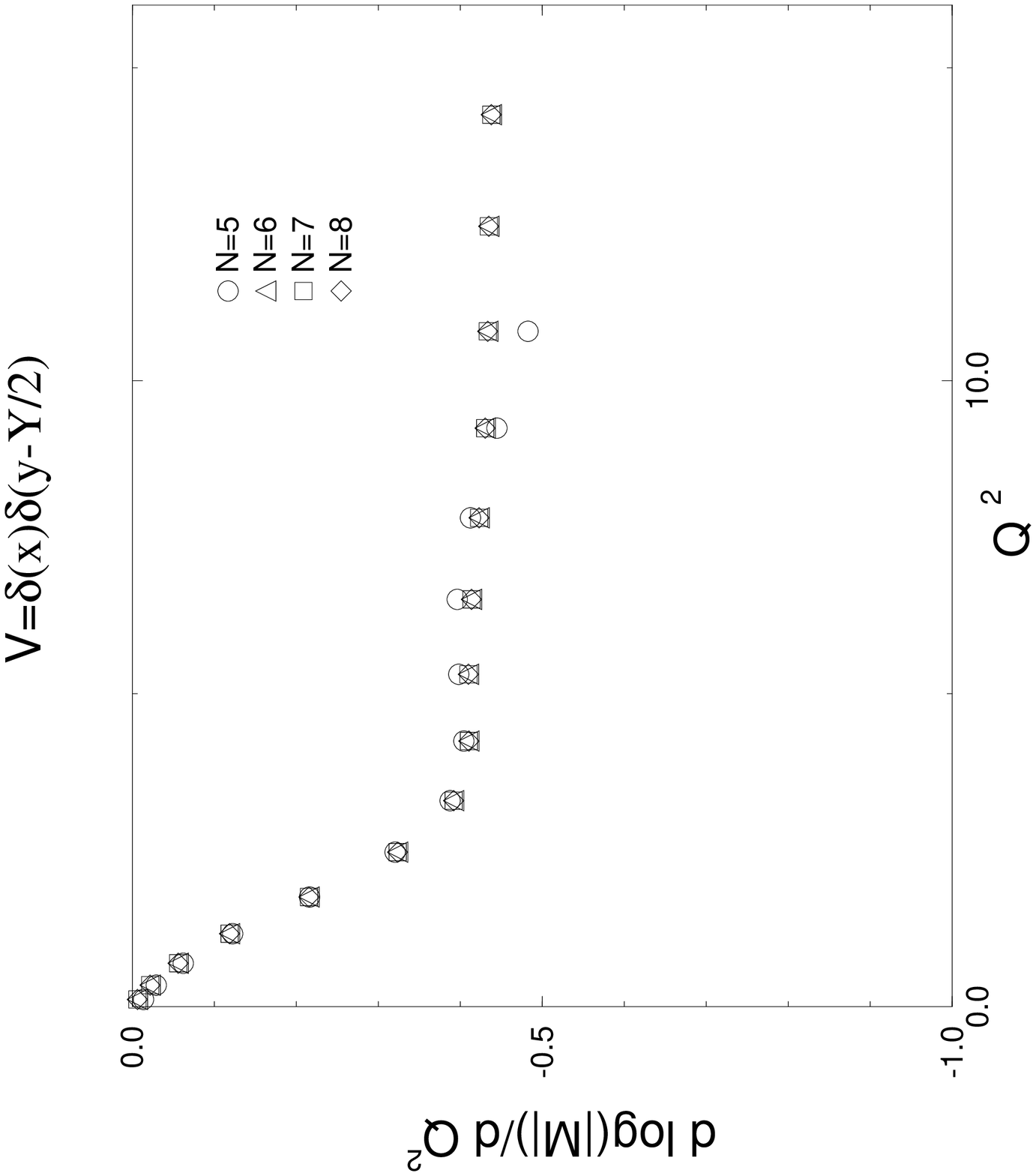,width=3.5in}}
\caption{
}
\vspace{0.5in} 
\end{figure}


\begin{references}
\bibitem{comm}  In Eq. (4.9) of ref. \protect\cite{TH}, the estimate
of the impurity matrix element  does not include an exponential suppression due to the many-body factor (the analogue of  $M(Q)$ of  Eq. (\protect\ref{MQ})  
for the torus).
\end{references}
\end{document}